\documentclass[reprint,amsmath,amssymb,aps]{revtex4-2}
\usepackage{amssymb}
\usepackage{subfigure}
\usepackage{xspace}
\usepackage{textpos}
\usepackage[english]{babel}
\usepackage{graphicx}
\usepackage{dcolumn}
\usepackage{bm}
\usepackage[colorlinks,linkcolor=blue,anchorcolor=blue,citecolor=blue,urlcolor=blue]{hyperref}
\usepackage[mathlines]{lineno}
\usepackage{overpic}
\usepackage{multirow}
\usepackage{color}
\usepackage{cancel}
\usepackage{ulem}
\newcommand{\jpsi}{J/\psi}
\newcommand{\pip}{\pi^+}
\newcommand{\pin}{\pi^-}

\newcommand{\pio}{\pi^0}

\begin{document}

\title{Measurements of the branching fractions of the $P$-wave charmonium spin-singlet state \texorpdfstring{$h_c(^1P_1) \to h^+ h^-\pio/\eta$}{}}

\author{
\begin{small}
\begin{center}
M.~Ablikim$^{1}$, M.~N.~Achasov$^{4,c}$, P.~Adlarson$^{76}$, O.~Afedulidis$^{3}$, X.~C.~Ai$^{81}$, R.~Aliberti$^{35}$, A.~Amoroso$^{75A,75C}$, Q.~An$^{72,58,a}$, Y.~Bai$^{57}$, O.~Bakina$^{36}$
% ...
\\
\vspace{0.2cm}
(BESIII Collaboration)\\
\vspace{0.2cm} {\it
$^{1}$ Institute of High Energy Physics, Beijing 100049, People's Republic of China\\
% ...
$^{3}$ Bochum  Ruhr-University, D-44780 Bochum, Germany\\
$^{4}$ Budker Institute of Nuclear Physics SB RAS (BINP), Novosibirsk 630090, Russia\\
% ...
$^{35}$ Johannes Gutenberg University of Mainz, Johann-Joachim-Becher-Weg 45, D-55099 Mainz, Germany\\
$^{36}$ Joint Institute for Nuclear Research, 141980 Dubna, Moscow region, Russia\\
% ...
$^{57}$ Southeast University, Nanjing 211100, People's Republic of China\\
$^{58}$ State Key Laboratory of Particle Detection and Electronics, Beijing 100049, Hefei 230026, People's Republic of China\\
% ...
$^{72}$ University of Science and Technology of China, Hefei 230026, People's Republic of China\\
% ...
$^{75}$ University of Turin and INFN, (A)University of Turin, I-10125, Turin, Italy; (B)University of Eastern Piedmont, I-15121, Alessandria, Italy; (C)INFN, I-10125, Turin, Italy\\
$^{76}$ Uppsala University, Box 516, SE-75120 Uppsala, Sweden\\
% ...
$^{81}$ Zhengzhou University, Zhengzhou 450001, People's Republic of China\\
\vspace{0.2cm}
$^{a}$ Deceased\\
% ...
$^{c}$ Also at the Novosibirsk State University, Novosibirsk, 630090, Russia\\
% ...
}\end{center}
\vspace{0.4cm}
\end{small}
}

\date{\today}

\begin{abstract}
Based on $(2712.4\pm 14.3)\times10^{6}$ $\psi(3686)$ events, we investigate four hadronic decay modes of the $P$-wave charmonium spin-singlet state $h_c(^1P_1) \to h^+ h^- \pi^0/\eta$ ($h=\pi$ or $K$) via the process $\psi(3686) \to \pi^{0}h_c$ at BESIII. 
  The $h_c \to \pi^+ \pi^- \pi^0$ decay is observed with a significance of 9.6$\sigma$ after taking into account systematic uncertainties. 
  Evidences for $h_c \to K^+ K^- \pi^0$ and $h_c \to K^+ K^- \eta$ are found with significances of $3.5\sigma$ and $3.3\sigma$, respectively, after considering the systematic uncertainties. 
  The branching fractions of these decays are measured to be $\mathcal{B}(h_c \to \pi^+ \pi^- \pi^0)=(1.36\pm0.16\pm0.14)\times10^{-3}$,  $\mathcal{B}(h_c \to K^+ K^- \pi^0)=(3.26\pm0.84\pm0.36)\times10^{-4}$, and $\mathcal{B}(h_c \to K^+ K^- \eta)=(3.13\pm1.08\pm0.38)\times10^{-4}$,
  where the first uncertainties are statistical and the second are systematic.
  No significant signal of $h_c\to\pi^+\pi^-\eta$ is found, and the upper limit of its decay branching fraction is determined to be $\mathcal{B}(h_c\to\pi^+\pi^-\eta) < 4.0 \times 10^{-4}$ at 90\% confidence level.
\end{abstract}

\maketitle


\section{Introduction}
The study of charmonium states plays a central role in our understanding of quantum chromodynamics~(QCD). 
Over the past years, the spin-singlet charmonium state $h_c(^1P_1)$ has been extensively studied, yet many of its decay modes are still unknown.
The first observation of $h_c$ was reported in 2005 by the CLEO experiment~\cite{PhysRevD.72.092004, PhysRevLett.95.102003}.
After that, the radiative decay $\it{h}_c \to \gamma\eta_c$ was confirmed by the Fermilab E835~\cite{PhysRevD.72.032001}, CLEO~\cite{PhysRevLett.101.182003}, and BESIII~\cite{PhysRevLett.104.132002} collaborations with an average branching fraction of (60$\pm$4)\%~\cite{workman2022}.
Recently, the BESIII collaboration reported the observation of several decay modes of $\it{h}_c \to$ \textit{light hadrons}, 
including $h_c\to p \bar{p}\pip\pin$~\cite{PhysRevD.99.072008},  $\it{h}_c \to \mathrm{2}(\pi^+ \pi^-)\pi^\mathrm{0}$~\cite{PhysRevD.99.072008},  $\it{h}_c \to K^+ K^- \pi^+ \pi^-\pi^\mathrm{0}$~\cite{PhysRevD.102.112007}, and $h_c \to 3(\pi^+ \pi^-)\pi^0$~\cite{search1_b}. 
The branching fractions are of the order of $10^{-3}$, and until now, the sum of the measured branching for $h_c$ decaying to light hadrons
is only 3-4~\%~\cite{workman2022}. 
From perturbative QCD (pQCD), the $\mathcal{B}(\it{h}_c \to light ~hadrons)$ is predicted to be about 48\%~\cite{PhysRevD.65.094024}, while a value of 8\% is obtained from non-relativistic QCD (NRQCD)~\cite{PhysRevD.65.094024}. 
This discrepancy between different theoretical models, as well between theory and the experimental measurements, motivates us to search for additional decay modes of the $h_c$ and to improve the measurement precision of the known $h_c$ decays with a larger $\psi(3686)$ data sample.

Although $h_c$ mesons cannot be produced directly in $e^+ e^-$ collisions at BESIII, the large $\psi(3686)$ sample with $(2712.4\pm 14.3)\times10^{6}$ $\psi(3686)$ events~\cite{psinum_b} provides an opportunity to study $h_c$ decays via the hadronic transition $\psi(3686) \to \pi^0 h_c$. 
In this paper, the four exclusive hadronic decays $h_c \to \pi^+ \pi^- \pi^0$, $h_c \to K^+ K^- \pi^0$, $h_c \to K^+ K^- \eta$, 
and $h_c\to\pi^+\pi^-\eta$, denoted as modes I, II, III, and IV, respectively,
are investigated.

\section{BESIII detector and Monte Carlo simulations}

The BESIII detector~\cite{ABLIKIM2010345} records symmetric $e^+e^-$ collisions 
provided by the BEPCII storage ring~\cite{Yu:IPAC2016-TUYA01}
in the center-of-mass energy range from 2.0 to 4.95~GeV,
with a peak luminosity of $1.1 \times 10^{33}\;\text{cm}^{-2}\text{s}^{-1}$ 
achieved at $\sqrt{s} = 3.773~\text{GeV}$. 
BESIII has collected large data samples in this energy region~\cite{Ablikim_2020, EcmsMea, EventFilter}. The cylindrical core of the BESIII detector covers 93\% of the full solid angle and consists of a helium-based
 multilayer drift chamber~(MDC), a plastic scintillator time-of-flight
system~(TOF), and a CsI(Tl) electromagnetic calorimeter~(EMC),
which are all enclosed in a superconducting solenoidal magnet
providing a 1.0~T magnetic field.
The solenoid is supported by an
octagonal flux-return yoke with resistive plate counter muon
identification modules (MUC) interleaved with steel. 
The charged-particle momentum resolution at $1~{\rm GeV}/c$ is
$0.5\%$, and the specific energy loss (${\rm d}E/{\rm d}x$)
resolution is $6\%$ for electrons
from Bhabha scattering. The EMC measures photon energies with a
resolution of $2.5\%$ ($5\%$) at $1$~GeV in the barrel (end cap)
region. The time resolution in the TOF barrel region is 68~ps, while
that in the end cap region was 110~ps. The end cap TOF
system was upgraded in 2015 using a multigap resistive plate chamber
technology, providing a time resolution of
60~ps,
which benefits 83\% of the data used in this analysis~\cite{etof1, etof22, etof3}.

Simulated data samples produced with a {\sc
geant4}-based~\cite{geant4} Monte Carlo (MC) package, which
includes the geometric description of the BESIII detector and the
detector response, are used to determine detection efficiencies
and to estimate backgrounds. The simulation models the beam
energy spread and initial-state radiation (ISR) in the $e^+e^-$
annihilations with the generator {\sc
kkmc}~\cite{kkmc, PhysRevD.63.113}. 
The inclusive MC sample includes the production of the
$\psi(3686)$ resonance, the initial-state radiation production of the $J/\psi$, and
the continuum processes incorporated in {\sc
kkmc}~\cite{kkmc, PhysRevD.63.113}.
All particle decays are modelled with {\sc
evtgen}~\cite{evtgen1,evtgen2} using branching fractions 
either taken from the
Particle Data Group (PDG)~\cite{workman2022}, when available,
or otherwise estimated with {\sc lundcharm}~\cite{PhysRevD.62.034, lundcharm}.
Final-state radiation from charged final-state particles is included using the {\sc photos} package~\cite{photos}.
The exclusive signal MC samples are generated by the phase space model with all the branching fractions of the intermediate states set to be 100\%, and each sample contains one hundred thousand events.

\section{Event selection and data analysis}
Charged tracks detected in the MDC are required to be within a polar angle ($\theta$) range of $|\!\cos\theta|<0.93$, where $\theta$ is defined with respect to the $z$ axis,
which is the symmetry axis of the MDC. 
For charged tracks, the distance of closest approach to the interaction point (IP) 
must be less than 10\,cm
along the $z$ axis,  
and less than 1\,cm
in the transverse plane.
The number of charged tracks is required to be two.
Particle identification~(PID) for charged tracks combines measurements of the d$E$/d$x$ in the MDC and the flight time in the TOF to form likelihoods $\mathcal{L}(h)~(h=p,K,\pi)$ for each hadron $h$ hypothesis.
For $h_c\to \pi^+\pi^-\pi^0$ and $h_c\to \pi^+\pi^-\eta$, the charged pions are required to satisfy $\mathcal{L}(\pi)>\mathcal{L}(K)$ and $\mathcal{L}(\pi)>\mathcal{L}(p)$, while for $h_c\to K^+ K^-\pi^0$ and $h_c\to K^+ K^-\eta$, the kaons are required to satisfy $\mathcal{L}(K)>\mathcal{L}(\pi)$ and $\mathcal{L}(K)>\mathcal{L}(p)$.

Photon candidates are identified using showers in the EMC.  The deposited energy of each shower must be more than 25~MeV in the barrel region ($|\!\cos \theta|< 0.80$) and more than 50~MeV in the end cap region ($0.86 <|\!\cos \theta|< 0.92$).
To exclude showers that originate from
charged tracks,
the angle subtended by the EMC shower and the position of the closest charged track at the EMC
must be greater than 10 degrees as measured from the IP. 
To suppress electronic noise and showers unrelated to the event, the difference between the EMC time and the event start time is required to be within 
[0, 700]\,ns.

The $\pio$ and $\eta$ candidates are reconstructed from $\gamma\gamma$ combinations with invariant mass windows in $(0.08, 0.20)$~${\rm GeV}/c^2$ and $(0.45, 0.65)$~${\rm GeV}/c^2$, respectively. 
The invariant mass of $\gamma\gamma$ is then constrained to the known mass of $\pio$ or $\eta$~\cite{workman2022} via a one-constraint (1C) kinematic fit requiring $\chi^2_{1{\rm C}}<200$.

To suppress background and improve the mass resolution,
a six-constraint (6C) kinematic fit, including the total initial four-momentum of the colliding beams, an invariant mass constraint of $\pio$ decaying from $\psi(3686)$, 
and an invariant mass constraint of $\pio$ or $\eta$ decaying from $h_c$, is performed according to the final states of each decay mode. 
The best $\pi^+ \pi^- (K^+ K^-)\pio\pio$ or $\pi^+ \pi^- (K^+ K^-)\pio\eta$ combination is selected with a minimum $\chi^2_{6{\rm C}}$ if there is more than one.
The requirement applied to $\chi^2_{6{\rm C}}$ is optimized using the Punzi figure of merit (FOM) $\frac{\epsilon}{a/2+\sqrt{B}}$~\cite{punzi}, 
where $\epsilon$ is signal efficiency, 
$a=5$ stands for the expected significance level, and $B$ represents the number of expected background events, estimated with the inclusive MC sample.

Considering the background events with three or five photons, the $\chi^2_{4{\rm C},n\gamma}~(n=3,~4,~5)$ values from a four-constraint (4C) kinematic fit together with three, four, and five photons are obtained.
We require $\chi^2_{4{\rm C},4\gamma}<\chi^2_{4{\rm C},3\gamma}$ and $\chi^2_{4{\rm C},4\gamma}<\chi^2_{4{\rm C},5\gamma}$ for 
modes II, III, and IV.
For mode I, 
only $\chi^2_{4{\rm C},4\gamma}<\chi^2_{4{\rm C},3\gamma}$ is applied to improve the ratio of signal to noise.
Another prominent background that originates from the decay of $\psi(3686)\to\pio\pio J/\psi$ or $\psi(3686)\to\eta J/\psi$ is vetoed by the requirement on the $\pio\pio$ or $\eta$ recoil mass to be outside the $J/\psi$ signal window.
When the final states include $\pio$, the momentum of $\pi^0$ from $\psi(3686)\to \pi^{0}\it{h}_c$ is usually lower than that from $h_c$ decay.
The former and latter $\pi^0$s are tagged as $\pi^0_{\rm L}$ and $\pi^0_{\rm H}$, respectively.
Furthermore, events containing resonances formed with the $\pio_{\rm L}$ are regarded as background events.
These background events are vetoed with additional selection criteria, especially for $\omega\to \pip\pin\pio$, $f_0(980)\to\pio_{\rm L} \pio_{\rm H}$, and $K^*(892)\to K\pio_{\rm L}$.
By combining one photon from $\pio$ decay and the other from $\eta$ decay, 
fake $\gamma\gamma$ combinations from $\pio$ decay are removed by the requirement on the invariant mass of $\gamma\gamma$
in the $h_c \to K^+ K^- \eta$ decay.
Finally, the background events from $\psi(3686)\to\pi^0\pi^0 J/\psi,~J/\psi\to\mu^+\mu^-$ with $\mu^+\mu^-$ misidentified as $\pi^+ \pi^-$ are vetoed by requiring a certain penetration depth in the MUC.
All detailed selection criteria are listed in Table~\ref{tab:table 1}. 
\begin{table}
\caption{\label{tab:table 1}%
Applied requirements on the $\chi^2_{6{\rm C}}$, invariant mass (M) windows, and recoil mass (RM) windows used as vetoes in each decay.
Here, $m$ denotes the known particle mass~\cite{workman2022}.
}
\begin{ruledtabular}
\begin{tabular}{lcc}
Mode &
$\chi^2_{6{\rm C}}$ &
Veto \\
\colrule
I &
$<30$ & $|{\rm RM}(\pio\pio)-m_{J/\psi}|>72~{\rm MeV}/c^2$\\
& & $|{\rm M}(\pip\pin\pio)-m_{\omega}|>28~{\rm MeV}/c^2$ \\ 
& & Depth of $\mu^{\pm} < 40~{\rm cm}$ \\
II &
$<50$ & $|{\rm RM}(\pio\pio)-m_{J/\psi}|>20~{\rm MeV}/c^2$ \\
& & $|{\rm M}(\pio_{\rm L}\pio_{\rm H})-m_{f_0(980)}|>110~{\rm MeV}/c^2$ \\
& & $|{\rm M}(K\pio_{\rm L})-m_{K^*(892)}|>40~{\rm MeV}/c^2$ \\
III &
$<35$ & $|{\rm RM}(\eta)-m_{J/\psi}|>10~{\rm MeV}/c^2$ \\
& & $|{\rm M}(\gamma\gamma)-m_{\pio}|>5~{\rm MeV}/c^2$ \\
& & $|{\rm M}(K\pio_{\rm L})-m_{K^*(892)}|>15~{\rm MeV}/c^2$ \\
IV &
$<24$ & $|{\rm RM}(\eta)-m_{J/\psi}|>10~{\rm MeV}/c^2$ \\
& & Depth of $\mu^{\pm} < 35~{\rm cm}$ \\

\end{tabular}
\end{ruledtabular}
\end{table}

After applying all the selection criteria, the remaining backgrounds from the inclusive MC sample are analyzed by the {\sc topoana}~\cite{topoan} package. 
These background events mostly have the same final states as the signals and are difficult to suppress. 
The background from the continuum production is estimated with the data sample taken at the center-of-mass energy of 3.650 GeV, with an integrated luminosity of 410 $\mathrm{pb}^{-1}$. 
The surviving events do not contribute to the peak position of $h_c$.

\section{Extraction of signal yields}
\begin{figure}
    \begin{overpic}[scale=.34]{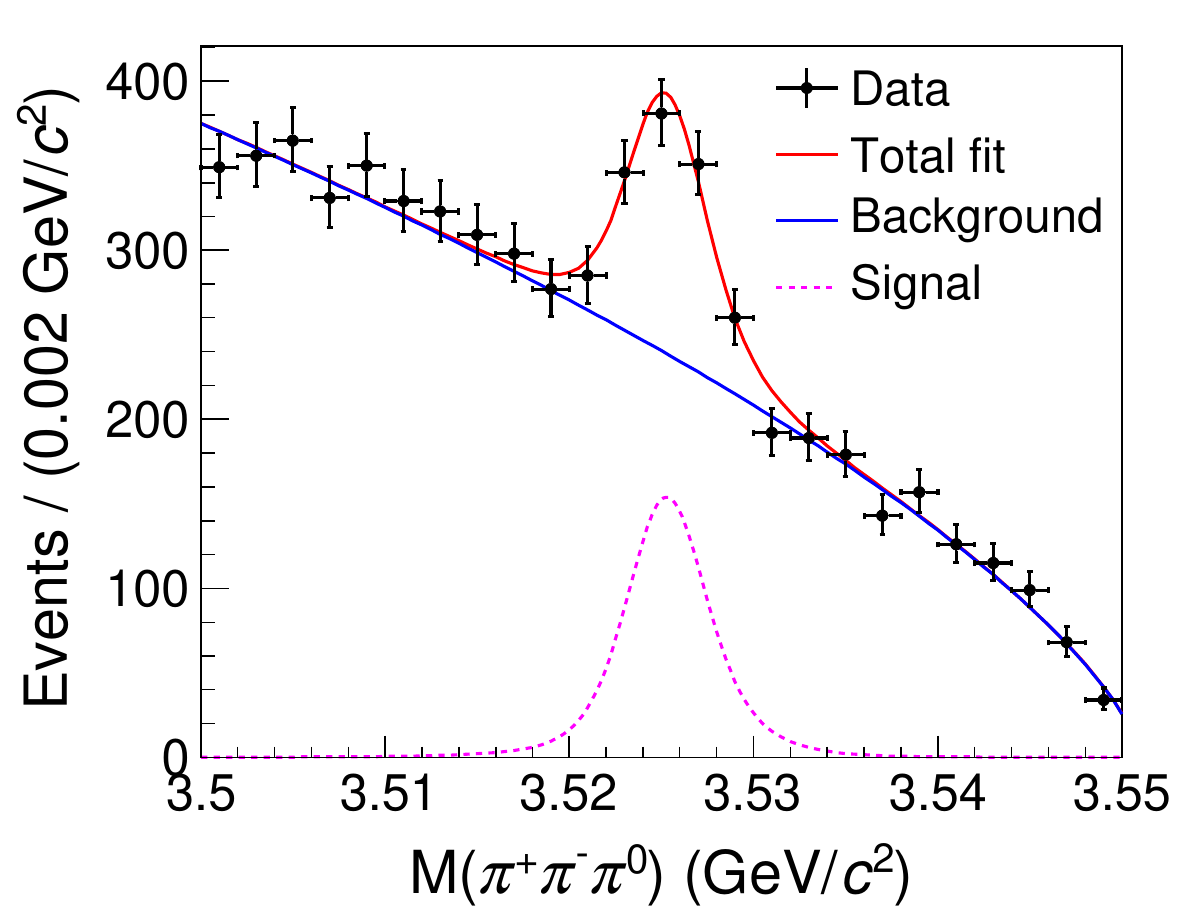}
    \put(30,28){\textbf{(a)}}
    \put(20,20){$\chi^2$/ndf $=0.84$}
    \end{overpic}
    \begin{overpic}[scale=.34]{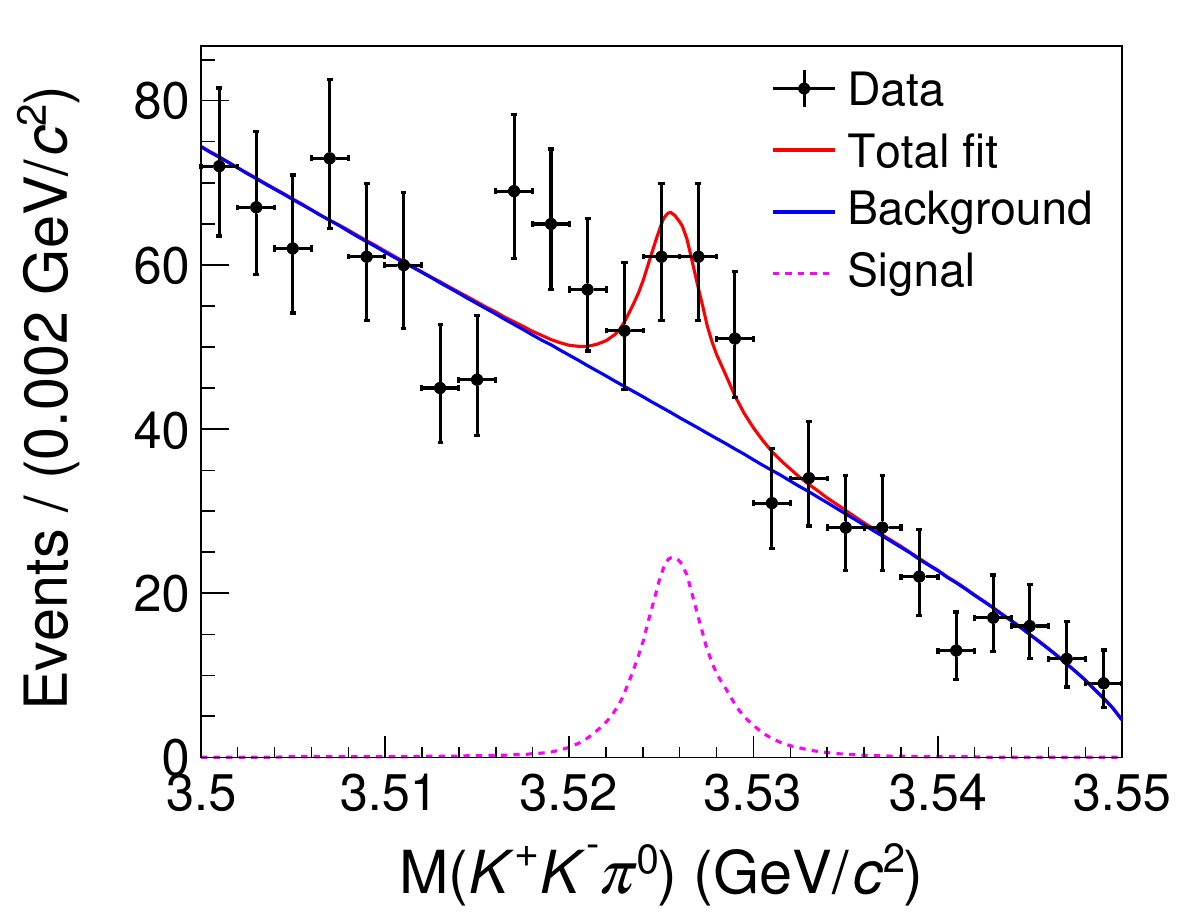}
    \put(30,28){\textbf{(b)}}
    \put(20,20){$\chi^2$/ndf $=1.02$}
    \end{overpic}
    \begin{overpic}[scale=.34]{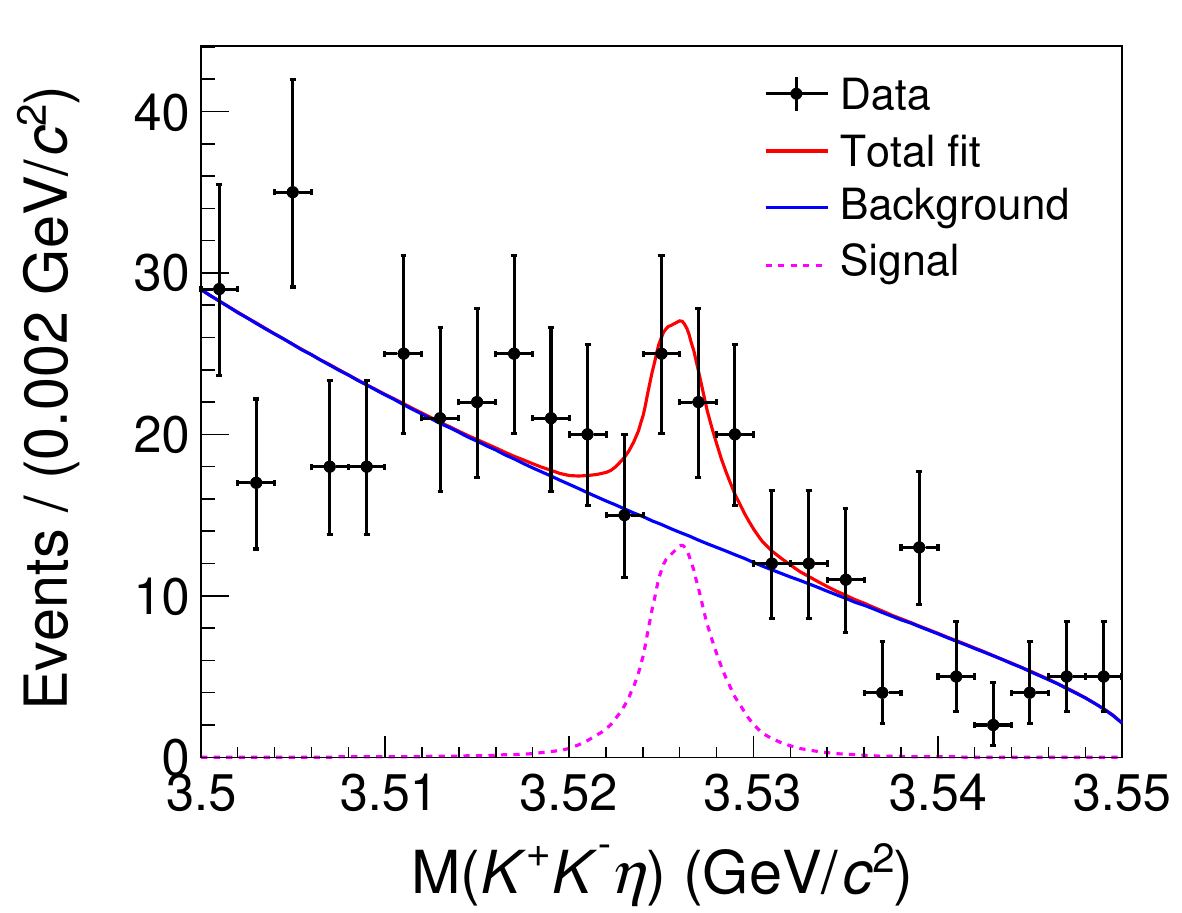}
    \put(30,28){\textbf{(c)}}
    \put(20,20){$\chi^2$/ndf $=1.15$}
    \end{overpic}
    \begin{overpic}[scale=.34]{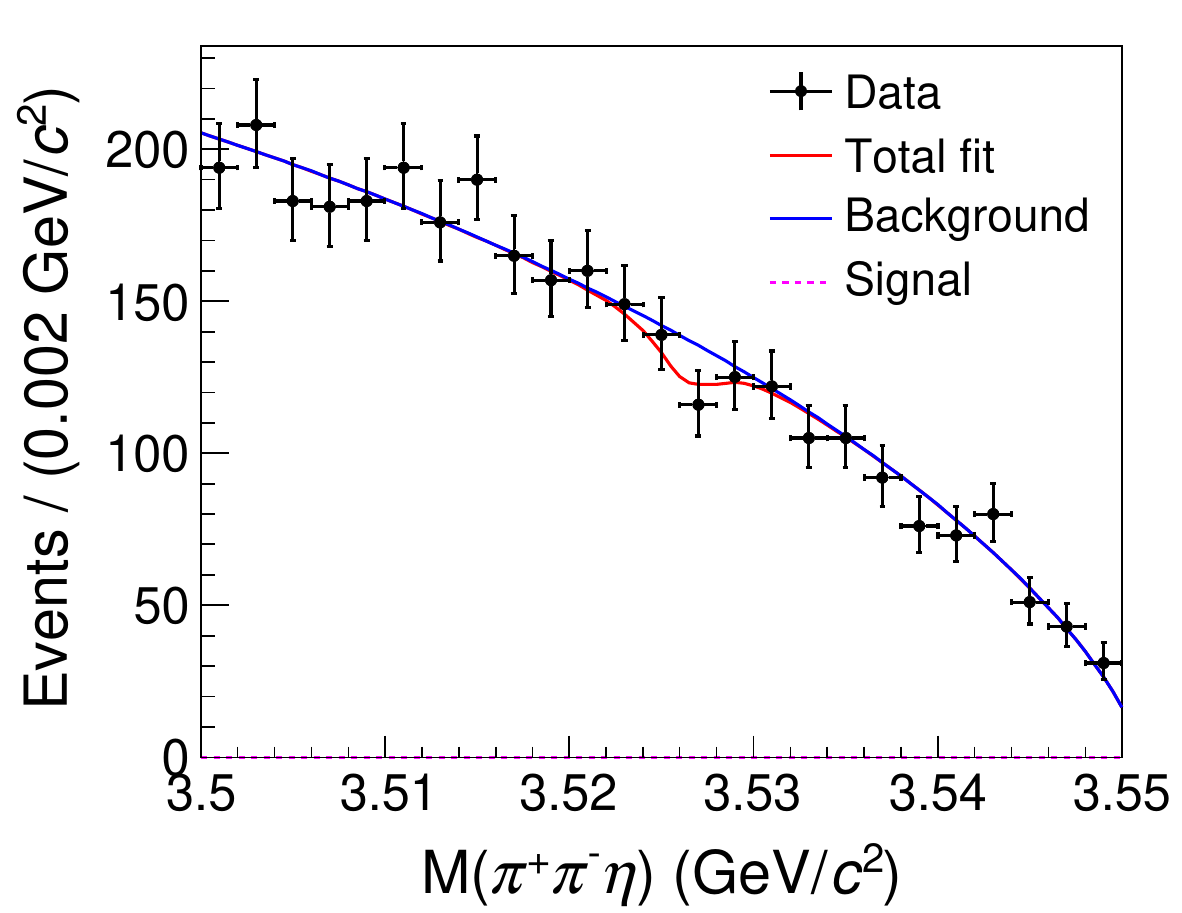}
    \put(30,28){\textbf{(d)}}
    \put(20,20){$\chi^2$/ndf $=0.74$}
    \end{overpic}    
        \caption{\label{fig:fit}The fits to the invariant mass distributions for (a) $h_c\to\pip\pin\pio$, (b) $h_c\to K^+ K^-\pio$, (c) $h_c\to K^+ K^-\eta$, and (d) $h_c\to\pip\pin\eta$.
        The black dots with error bars are the data, the red solid lines represent the fit results, the blue solid lines indicate the backgrounds, and the pink dashed lines illustrate the signals.}
    \end{figure}
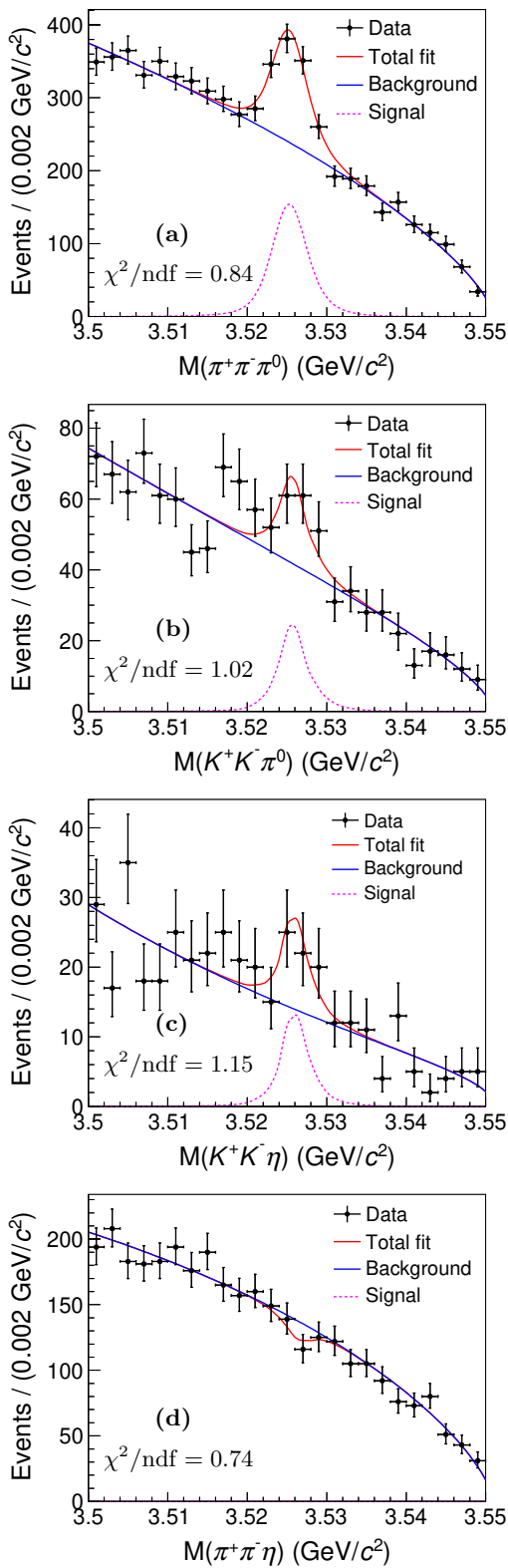

To determine the number of signal events $N_{\rm sig}$ in each decay, unbinned maximum likelihood fits are performed to the invariant mass spectra as shown in Fig~\ref{fig:fit}.
In the fit, the $h_c$ signal is described by an MC-simulated shape convolved with a free-parameter Gaussian function accounting for the mass resolution difference between data and MC simulation. 
The mass resolution estimated from MC simulation is around 1.2 MeV$/c^2$ for each $h_c$ decay channel.
The background shape is represented by an ARGUS function~\cite{argus}, 
with the end point set to the kinematic threshold of 3.551 GeV/$c^2$.

The branching fraction of each signal decay is determined by
\begin{equation} \label{eq:1}
     \mathcal{B}(h_c\to h^+ h^-\pio/\eta)=
     \frac{N_{\rm sig}}{N_{\psi(3686)}\cdot\prod_i\mathcal{B}_i\cdot\epsilon},   
\end{equation}
where
$N_{\rm sig}$ stands for the number of signal events obtained from the fit; $N_{\psi(3686)}$ represents the total number of $\psi(3686)$ events~\cite{psinum_b}, 
$\prod_i\mathcal{B}_i=\mathcal{B}(\psi(3686)\to\pio h_c)\cdot\mathcal{B}(\pio\to\gamma\gamma)\cdot\mathcal{B}(\pio/\eta\to\gamma\gamma)$, 
where $\mathcal{B}(\psi(3686)\to\pio h_c)$ is the branching fraction of $\psi(3686)\to\pio h_c$, 
and $\mathcal{B}(\pio\to\gamma\gamma)\cdot\mathcal{B}(\pio/\eta\to\gamma\gamma)$ is the product of the branching fractions.
These branching fractions
are quoted from the PDG~\cite{workman2022}. 

\begin{figure*}[htbp]
	\mbox{
		\begin{overpic}
			[scale=0.28]
			{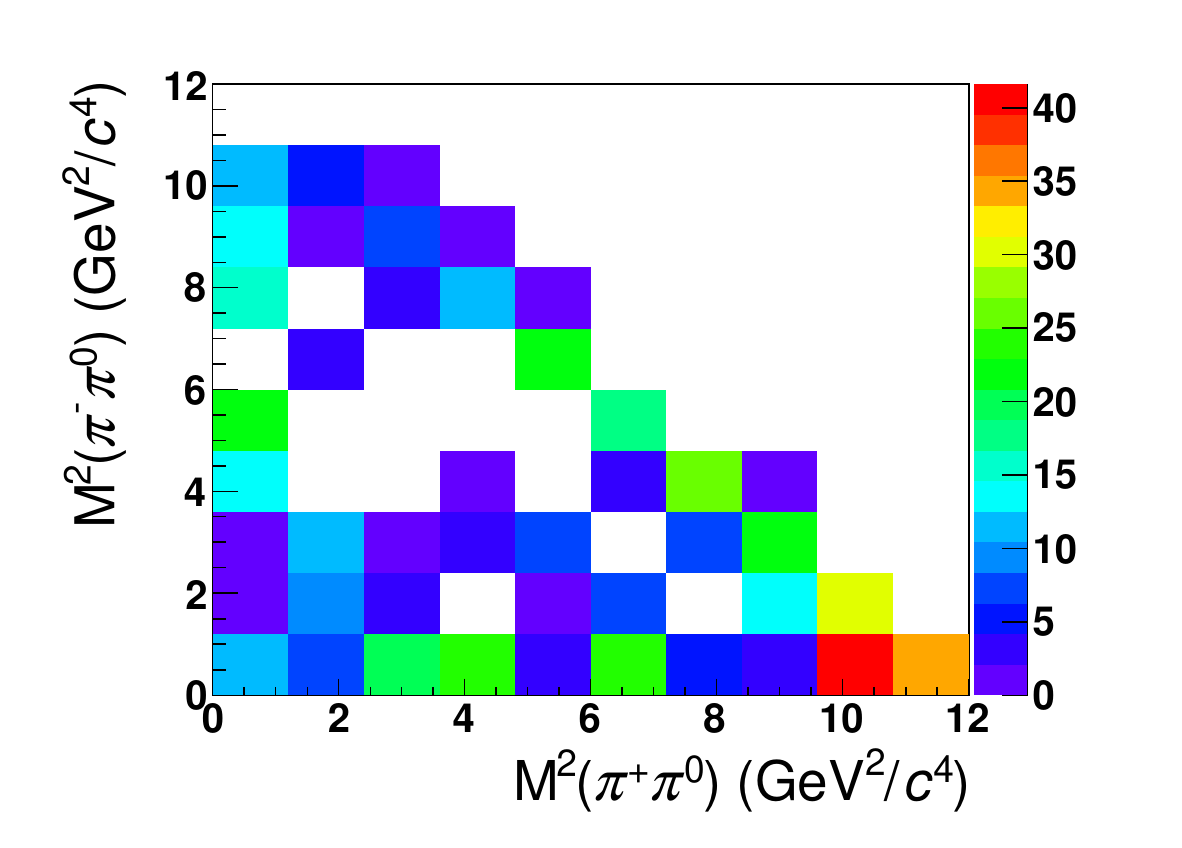}
			\put(65,55){\textbf{(a)}}
		\end{overpic}
		\begin{overpic}
			[scale=0.28]
			{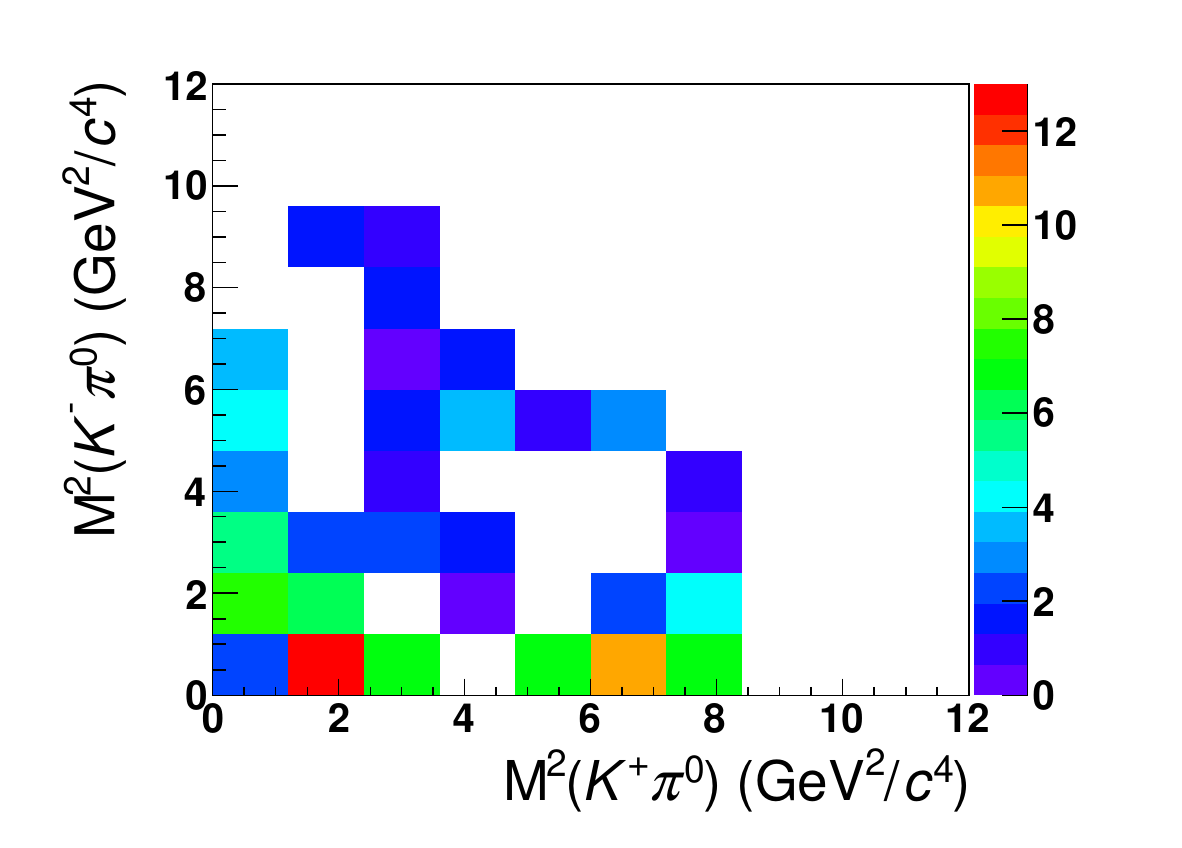}
			\put(65,55){\textbf{(b)}}
		\end{overpic}
		\begin{overpic}
			[scale=0.28]
			{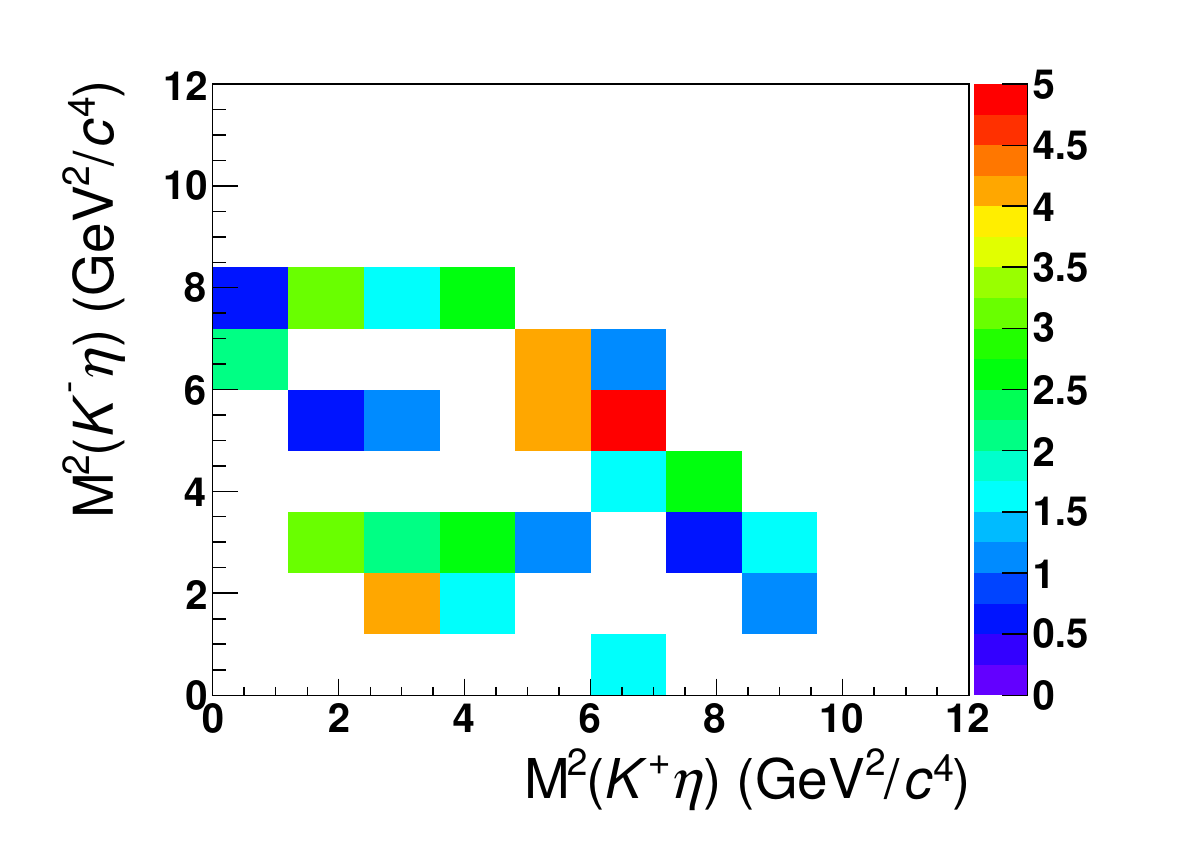}
			\put(65,55){\textbf{(c)}}
		\end{overpic}
	}
	\caption{Dalitz plots of data with the background events subtracted for (a) $h_c\to\pip\pin\pio$, (b) $h_c\to K^+ K^-\pio$, and (c) $h_c\to K^+ K^- \eta$.}
	\label{fig:2}
\end{figure*}

For each signal decay, we obtain the Dalitz plot of ${\rm M}^2(h^+\pio/\eta)$ versus ${\rm M}^2(h^-\pio/\eta)$ in the $h_c\to h^+ h^-\pio/\eta$ signal region defined as $3.52<{\rm M}(h^+ h^-\pio/\eta)<3.53$ GeV/$c^2$ from data, 
where the background events have been subtracted using the normalized $h_c$ sidebands defined as $(3.505, 3.515)$ GeV/$c^2$ or $(3.535, 3.545)$ GeV/$c^2$.
The corresponding Dalitz plots are shown  in Fig.~\ref{fig:2},
where we divide the Dalitz plot of data into $10\times10$ bins, with a bin size of 1.2~${\rm GeV}^2$/$c^4$.
Due to limited statistics, no intermediate resonances are found.
The detection efficiencies are obtained from signal MC simulation, corrected by the Dalitz plots observed in data for modes I, II, and III as follows.
The corrected efficiencies are determined via
\begin{equation} \label{eq:5.6}
    \epsilon=\frac{\sum_i s_i \cdot \epsilon_i}{S},
\end{equation}
where $i$ runs over all bins, $s_i$ and $\epsilon_i$ are the number of signal candidates in data and the efficiency from MC events for the $i$-th bin, and
$S$ is the total number of $h_c$ signal candidates in data. 
The efficiencies are 17.7\%, 9.7\%, 13.1\%, and 14.2\% for modes I, II, III, and IV, respectively.

The statistical significance is estimated from the difference of the logarithmic likelihoods of the fits without and with a signal component
with the difference in the number of degrees of freedom ($\Delta_{\rm ndf}=3$) that are considered.
The significance of 
mode I
is $10.8\sigma$ and the evidences for the decay modes 
II and III
are found with statistical significances of $3.5\sigma$ and $3.3\sigma$, respectively.
After considering systematic effects, the resulting significances are 9.6$\sigma$, 3.5$\sigma$, and 3.3$\sigma$ for 
modes I, II, and III
respectively.
Since no obvious signal of 
mode IV
is observed,
assuming that the number of signal events follows a Poisson distribution with a uniform prior probability density function,
a Bayesian upper limit~\cite{PhysRevD.57.3873} of 
mode IV
is determined. 

\begin{table*}
\caption{\label{tab:table2}%
The branching fractions (or their upper limit at 90\% confidence level) for each $h_c$ decay together with the corrected detection efficiencies, the fitted signal yields, and the signal significances with systematic uncertainties included. The first uncertainties are statistical, the second are systematic.}
\begin{ruledtabular}
\begin{tabular}{lccccc}
 Mode
 & $\epsilon~(\%)$ & Signal yield & Signal significance & $\mathcal{B}$~[This work]& $\mathcal{B}$ [PDG]~\cite{workman2022} \\ \hline
 I
 & 17.7 & $472\pm56$ & 9.6$\sigma$ & $(1.36\pm0.16\pm0.14)\times10^{-3}$ &$(1.9\pm0.5)\times10^{-3}$\\
 II
 &9.7 & $62\pm16$ & 3.5$\sigma$ & $(3.26\pm0.84\pm0.36)\times10^{-4}$ &$<6\times10^{-4}$ \\
 III
 &13.1 & $32\pm11$ & 3.3$\sigma$ & $(3.13\pm1.08\pm0.38)\times10^{-4}$ &$<1.0\times10^{-3}$\\
 IV
 & 14.2 & $<44.5$ &$\cdots$ & $<4.0\times10^{-4}$&$\cdots$ \\

\end{tabular}
\end{ruledtabular}
\end{table*}

\section{Systematic uncertainties}
In the branching fraction measurements, the systematic uncertainties are divided into multiplicative and additive terms.
The multiplicative terms include tracking, PID, $\pi^0$ and $\eta$ reconstruction, kinematic fit, selection criteria, number of $\psi(3686)$ events, and quoted branching fractions. 
The additive terms originate from fit range, signal shape, and background shape in the fit procedure.

The tracking efficiency is estimated with the control samples of $\jpsi\to p\Bar{p}\pip\pin$~\cite{Yuan-2016} and $e^+ e^-\to\pip\pin K^+ K^-$~\cite{PhysRevD.99.091103}. 
The resulting systematic uncertainty due to the tracking is assigned to be 1.0\% for each charged pion or kaon.
The uncertainties due to the PID for charged pion and kaon are studied with the control samples of $J/\psi\to\pip\pin\pio$ and $J/\psi\to K^+ K^-\pio$~\cite{PhysRevD.83.112005}, and are assigned to be 1.0\% for each pion or kaon.

The systematic uncertainty of the $\pi^0$ reconstruction is estimated using the control samples of $\psi(3686)\to \pi^0 \pi^0 J/\psi$ and $e^+ e^-\to \omega \pi^0$, resulting in the uncertainty as a function of $\pi^0$ momentum. 
The systematic uncertainties in the $\pi^0$ reconstruction are assigned according to the momentum of $\pi^0$ in the final state, which are 5.5\%, 5.2\%, 1.0\%, and 1.0\% for the decay modes I, II, III, and IV, respectively.
The systematic uncertainty of the $\eta$ reconstruction is estimated to be 1.0 \% using a high purity control sample of $\jpsi\to p\Bar{p}\eta$~\cite{PhysRevD.81.052005}.
Since the uncertainties between two $\pi^0$s or between $\pi^0$ and $\eta$ are assumed to be correlated, their values are added linearly.

The systematic uncertainties associated with the 6C kinematic fit are assigned as the differences between the efficiencies before and after the helix correction~\cite{PhysRevD.87.012002},
which are 0.6\%, 1.0\%, 0.9\%, and 0.9\% for modes I, II, III, and IV, respectively.

A Barlow test~\cite{barlow2002systematic} is performed to evaluate the uncertainty from the mass window requirements.
The significant deviation $\zeta$ is defined as
    \begin{equation}\label{eq:3}
        \zeta=
        \frac{|\mathcal{B}_{\rm nominal}-\mathcal{B}_{\rm test}|}
        {\sqrt{|\sigma^2_{\mathcal{B}_{\rm nominal}}
        -\sigma^2_{\mathcal{B}_{\rm test}}|}},
    \end{equation}
    where $\mathcal{B}$ represents the branching fractions and $\sigma^2_{\mathcal{B}}$ is the statistical uncertainty of $\mathcal{B}$.
To obtain the $\zeta$ distributions, 
we examine the branching fractions after enlarging or shrinking the veto region. 
For different background vetoes, we vary the corresponding mass windows several times with a step of 1 MeV/$c^2$.
Since there is no obvious change larger than $1.0\sigma$ for each requirement, we do not assign a systematic uncertainty to this term.

The uncertainty of the total number of $\psi(3686)$ events in data is 0.5\%~\cite{psinum_b}.
In the branching fraction calculations, the $\mathcal{B}(\psi(3686) \to \pi^0 h_c)$, $\mathcal{B}(\pi^0\to\gamma\gamma)$, and $\mathcal{B}(\eta\to\gamma\gamma)$ are quoted from PDG. 
The total uncertainty due to these quoted branching fractions, which is dominated by $\mathcal{B}(\psi(3686) \to \pi^0 h_c)$, is 5.5\%~\cite{workman2022, PhysRevD.106.072007}.

The additive systematic uncertainties are estimated below. The uncertainties due to the fit range are estimated by enlarging or shrinking the nominal fit range of (3.50, 3.55)~GeV$/c^2$ to be (3.49, 3.55)~GeV$/c^2$ and (3.51, 3.55)~GeV$/c^2$. 
The differences of the fitted signal yields between the nominal and alternative results are taken as the systematic uncertainties.
In the nominal fits, the signals are described by the MC-simulated shape convolved with a free-parameter Gaussian function. 
Since the numbers of signal events are limited for the decay modes II, III, and IV, the uncertainty from the mass resolution difference between data and MC simulation is estimated by changing the standard deviation of the Gaussian function to 1~MeV/$c^2$, which is determined from the $h_c\to\pi^+\pi^-\pi^0$ model. 
The differences relative to the nominal fits are taken as the systematic uncertainties.
The uncertainties due to the background shape are considered by changing the nominal ARGUS background shape to a second-order Chebyshev polynomial function.
The differences in the fitted signal yields are taken as the systematic uncertainties.

When the upper limit of the number of signal events in 
mode IV
is determined,
the additive systematic uncertainties are considered by retaining the largest upper limit by a maximum-likelihood fit with different fit ranges, signal shapes, and background shapes.
To incorporate the multiplicative terms, 
we convolve the likelihood distribution $\mathcal{L}(N_{\rm sig})$ with the quadratic sum of the multiplicative terms~\cite{stenson2006exact,Liu_2015}.

The individual uncertainties are assumed to be independent and added in quadrature to obtain the total systematic uncertainty.
In Table~\ref{tab:15}, the multiplicative terms are listed separately, and ``Fit procedure'' denotes the quadratic sum of the additive terms.
\begin{table}[htbp]
\caption{\label{tab:15} The relative systematic uncertainties~(\%) on the branching fraction measurements. 
In mode IV, there are no systematic uncertainties of the efficiency correction and fit procedure due to no evidence for a signal in the data.
The uncertainty due to the fit procedure is additive, the others are multiplicative.}
    \begin{tabular}{lcccc} \hline 
        Source/Mode&I & II & III & IV\\ \hline

         Tracking & 2.0 & 2.0 & 2.0 & 2.0\\
         PID & 2.0 & 2.0 &2.0& 2.0\\
         Reconstruction of $\pio$ and $\eta$ & 5.5 & 5.2 & 2.0 & 2.0\\
         Kinematic fit& 0.6 & 1.0 & 0.9 & 0.9\\
         $N_{\psi(3686)}$& 0.5 & 0.5 & 0.5 &0.5\\
         $\mathcal{B}(\psi(3686)\to\pi^0 h_c)$& 5.5 & 5.5 & 5.5 & 5.5\\ 
         Fit procedure& 6.6  & 7.3 & 10.1& $\cdots$\\\hline
         Sum & 10.6 & 10.9 & 12.1 & 6.6 \\ \hline
    \end{tabular}
\end{table}

\section{Summary}
Using $(2712.4\pm 14.3)\times10^{6}$ $\psi(3686)$ events, the modes
$h_c \to \pi^+ \pi^- \pi^0$, $h_c \to K^+ K^- \pi^0$, and $h_c \to K^+ K^- \eta$
are found with significances of 9.6$\sigma$, 3.5$\sigma$, and 3.3$\sigma$, respectively.
Their decay branching fractions are determined to be $\mathcal{B}(h_c \to \pi^+ \pi^- \pi^0)=(1.36\pm0.16\pm0.14)\times10^{-3}$,  $\mathcal{B}(h_c \to K^+ K^- \pi^0)=(3.26\pm0.84\pm0.36)\times10^{-4}$, and $\mathcal{B}(h_c \to K^+ K^- \eta)=(3.13\pm1.08\pm0.38)\times10^{-4}$, where the first uncertainties are statistical and the second are systematic. 
These results are consistent with the PDG values~\cite{workman2022}.
No obvious signal of 
$h_c\to\pi^+\pi^-\eta$ decay
is observed, and the upper limit of its decay branching fraction  
is determined to be $\mathcal{B}(h_c\to\pi^+\pi^-\eta) < 4.0\times$ $10^{-4}$ at 90\% confidence level.
The isospin conservation in the process $h_c\to K \Bar{K}\pi$ leads to a relationship of $\mathcal{B}(h_c\to K^+ K^-\pi^0):\mathcal{B}(h_c\to K^0 K^-\pi^+):\mathcal{B}(h_c\to \Bar{K}^0 K^+\pi^-):\mathcal{B}(h_c\to K^0 \Bar{K}^0 \pi^0)=1:2:2:1$. 
The $\mathcal{B}(h_c\to K \Bar{K}\pi)=(0.20\pm0.05)\%$ measured in this work is consistent with that obtained in $h_c\to K^0_S K^+\pi^-$ decay mode~\cite{kskpi_b}.
Theoretically, $\mathcal{B}(h_c\to K \Bar{K}\pi)$ is predicted to be $(1.4\pm0.9)\%$ for pQCD or $(5.5\pm3.3)\%$ for NRQCD~\cite{PhysRevD.65.094024}.
Since the large uncertainty mainly arises from the input values of $\mathcal{B}(\eta_{c}\to K\bar{K}\pi$) and $\Gamma(\eta_{c})$ quoted from PDG 2000~\cite{pdg2000}, 
we take the most recent results from the PDG~\cite{workman2022} and reweigh the $\mathcal{B}(h_c\to K \Bar{K}\pi)$ in Ref.~\cite{PhysRevD.65.094024}.
The branching fraction of $h_c\to K\bar{K}\pi$ is calculated to be $(4.4\pm0.8)\%$ with pQCD or $(16.9\pm7.3)\%$ with NRQCD,
 which strongly deviates from the experimental results. 
Since only the leading-order terms are evaluated in the theoretical calculations from both pQCD and NRQCD, 
higher precision from theory is desirable.

\acknowledgements
The BESIII Collaboration thanks the staff of BEPCII and the IHEP computing center for their strong support. This work is supported in part by National Key R\&D Program of China under Contracts Nos. 2020YFA0406300, 2020YFA0406400, 2023YFA1606000; National Natural Science Foundation of China (NSFC) under Contracts Nos. 11635010, 11735014, 11935015, 11935016, 11935018, 12025502, 12035009, 12035013, 12061131003, 12192260, 12192261, 12192262, 12192263, 12192264, 12192265, 12221005, 12225509, 12235017, 12361141819; the Chinese Academy of Sciences (CAS) Large-Scale Scientific Facility Program; the CAS Center for Excellence in Particle Physics (CCEPP); Joint Large-Scale Scientific Facility Funds of the NSFC and CAS under Contract No. U2032108; 100 Talents Program of CAS; The Institute of Nuclear and Particle Physics (INPAC) and Shanghai Key Laboratory for Particle Physics and Cosmology; German Research Foundation DFG under Contracts Nos. 455635585, FOR5327, GRK 2149; Istituto Nazionale di Fisica Nucleare, Italy; Ministry of Development of Turkey under Contract No. DPT2006K-120470; National Research Foundation of Korea under Contract No. NRF-2022R1A2C1092335; National Science and Technology fund of Mongolia; National Science Research and Innovation Fund (NSRF) via the Program Management Unit for Human Resources \& Institutional Development, Research and Innovation of Thailand under Contract No. B16F640076; Polish National Science Centre under Contract No. 2019/35/O/ST2/02907; The Swedish Research Council; U. S. Department of Energy under Contract No. DE-FG02-05ER41374.



\end{document}